\documentclass{article}

\usepackage[english]{babel}

\usepackage[letterpaper,top=2cm,bottom=2cm,left=3cm,right=3cm,marginparwidth=1.75cm]{geometry}

\usepackage{amsmath}
\usepackage{graphicx}
\usepackage[colorlinks=true, allcolors=blue]{hyperref}
\usepackage{authblk}

\title{Modeling Methane Intensity of Oil and Gas Upstream Activities by Production Profile}
\author[1]{Quentin Peyle}
\author[2]{Imène Ben Rejeb-Mzah}
\author[1,3]{Baptiste Piofret}
\author[1]{Antoine Benoit}
\author[1,4]{Alexandre d'Aspremont}
\author[2]{Adil El Yaalaoui}
\affil[1]{Kayrros SAS}
\affil[2]{BNP Paribas}
\affil[3]{Accuracy}
\affil[4]{CNRS - Ecole Normale Supérieure}

\begin{document}
\maketitle

\begin{abstract}
We propose a methodology for modelling methane intensities of Oil and Gas upstream activities for different production profiles with diverse combinations of region of operation and production volumes associated. This methodology leverages different data sources, including satellite measurements and public estimates of methane emissions but also country-level oil and gas production data and company reporting. The obtained methane intensity models are compared to the reference companies’ own reporting in order to better understand methane emissions for different types of companies. The results show that regions of operation within the different production profiles have a significant impact on the value of modelled methane intensities, especially for operators located in a single or few countries, such as national and medium-sized international operators. This paper also shows that methane intensities reported by the companies tend to be on average 16.1 times smaller than that obtained using the methodology presented here, and cannot account for total methane emissions that are estimated for upstream operations in the different regions observed.
\end{abstract}

\section{Introduction}

\subsection{Context}

Oil and natural gas are major industries in the energy market and play an influential role in the global economy as the world's primary fuel source. They cover more than half of the planet’s demand for primary energy, supplying the world with 28 billion barrels of oil and 25 billion barrels of oil equivalent of gas in 2020. The industry is facing increasing demands to clarify the implications of the energy transition for their operations and business models, and to explain the contribution that they can make to reducing greenhouse gas emissions in order to achieve the goals of the Paris Agreement.

As of today, the oil and gas industry is one of the most important contributors to anthropogenic methane emissions, representing 60\% of global methane emissions related to the operations of the energy sector or 80 megatons of CH4, the remaining 40\% coming from coal and bio-energy. According to the International Energy Agency (``IEA"), emissions from the industry must be cut in half by 2030 to meet global energy and climate goals. The reduction of accidental or maintenance-related methane releases has been identified as the most important and cost-effective way for the industry to reduce its emissions.

\subsection{Use case in the banking industry}

In order to implement its commitment to fight climate change, the banking industry launched several initiatives like the commitment signed by five international banks (BNP Paribas, SG, ING, BBVA and SC) during COP 24 in Katowice, to develop open-source methods and tools for measuring the alignment of lending portfolios with the goals of the Paris Agreement (COP 21). What’s more, these banks aspired to then lead the
implementation of these methodologies and tools to actually align their lending portfolios with these climate goals. After that, the Net-Zero Banking Alliance (NZBA) was launched on 21 April 2021 under the umbrella of the United Nations Environment Programme Finance Initiative (UNEP FI). This banking alliance is a decisive step in the mobilisation of the financial sector for climate as its members committed to align their financing on net zero carbon trajectories by 2050 or sooner which is consistent with the goal of limiting the average global temperature to 1.5° above pre-industrial levels.

The IPCC specified in its Sixth Assessment Report (AR6) \cite{AR6} that all global modelled pathways that limit warming to 1.5°C with no or limited overshoot, and those that limit warming to 2°C, involve rapid and deep and, in most cases, immediate greenhouse gas emissions reductions in all sectors this decade. These pathways require for instance the reduction of global methane emissions by 34\% from 2019 level by 2030. Hence, a
specific focus should be placed by NZBA banks on their financed methane emissions reduction.

When they signed the NZBA statement, NZBA banks pledged to set sector-specific targets covering at least ten of the most emitting sectors including the O\&G sector \cite{NZBA}. Knowing that around 70\% of methane emissions from fossil fuel operations could be reduced with existing technology according to the IEA \cite{IEA}, one of the potential use cases by the banking industry of this study is to improve client level methane emissions measurement and as a consequence to improve the portfolio level methane footprint measurement. This would allow for better  monitoring of the portfolio methane emissions and contribute to the implementation of methane risk mitigation actions. This would also contribute to a better assessment of the environmental performance and risk of O\&G industry players.

\section{Literature Review}

In the pursuit of understanding and mitigating methane emissions, a growing body of literature highlights the considerable underestimation of reported emissions from oil and gas activities, revealing flaws in existing inventories.

The study conducted by Stuart N. Riddick et al. (2022) \cite{literatureReview6} for the Royal Society of Chemistry revealed a substantial discrepancy in the United Kingdom's National Atmospheric Emissions Inventory (NAEI) estimates for methane emissions from oil and natural gas extraction and transport. The NAEI's bottom-up approach was found to be five times lower than the alternative integrated approaches proposed in the study, which combined direct measurements, management practices, and environmental conditions. The underestimation was attributed to outdated or incorrect emission factors, incomplete activity data, and insufficient information on vented emissions, indicating a need for more accurate and comprehensive methodologies. Another study focusing on the Permian Basin in the United States, published by Jevan Yu et al. (2022) \cite{literatureReview5} from the Environmental Defense Fund, brought attention to the underestimation of methane emissions from natural gas gathering pipelines. The study, utilizing aerial campaigns and demonstrating the limitations of ground-based monitoring, revealed emission factors 14–52 times higher than the U.S. Environmental Protection Agency's national estimate for gathering lines. 

Yet for a long time, abatement efforts were hampered by overwhelming detection and measurement difficulties. In this context, the scientific community has delved into recent advances in satellite-based technology, revealing its transformative potential in estimating methane emission inventories at both local and global levels. The following studies demonstrate the crucial role of satellites and align with previous research highlighting the disparity between measured and reported inventories.
Lu Shen et al. (2021) \cite{fullInversion2} harnessed data from the Tropospheric Monitoring Instrument (TROPOMI) to scrutinize methane emissions in eastern Mexico. Their findings illuminated substantial disparities between satellite-derived estimates and national inventories, with a pronounced focus on emissions within the oil and gas sector. Ramon Alvarez et al. (2018) \cite{literatureReview1} conducted a comprehensive assessment of methane emissions within the U.S. oil and gas supply chain, highlighting the utility of satellite data in revealing previously underestimated emissions, particularly during abnormal operating conditions. Expanding on this trajectory, Lu Shen et al. (2022) \cite{literatureReview2} leveraged TROPOMI to quantify methane emissions from individual oil and gas basins in the United States and Canada. Once again, satellite technology illuminated emissions discrepancies, both at the national and regional levels. Extending our perspective beyond North America, Zichong Chen et al. (2023) \cite{literatureReview3} employed TROPOMI satellite observations to analyze methane emissions in the Middle East and North Africa. This research demonstrated the potential for TROPOMI to optimize and separate national emissions by sector, revealing that regional anthropogenic emissions exceeded prior inventories, notably in the oil and gas sector. Finally, a study published by Lu Shen et al. (2023) \cite{literatureReview4} in Nature used high-resolution inversions of satellite observations to globally quantify methane emissions from fossil fuel exploitation. The research exposed a 30\% underreporting of oil and gas emissions in national inventories submitted to the United Nations Framework Convention on Climate Change (UNFCCC), emphasizing the global scale of the discrepancy. 

These studies collectively underscore the transformative potential of satellite-based methane monitoring, offering global perspectives that can expose challenging emissions sources and discrepancies between reported inventories and actual emissions. This technological innovation equips us with vital data to make informed decisions and drive proactive measures in our ongoing battle against methane emissions, representing a significant stride towards a more sustainable and responsible future.

\section{Methodology}

Methane emissions from upstream oil and gas operations include releases of methane related to crude oil or natural gas operations in producing basins. Methane emissions typically come from wells, first-mile pipeline compressor stations, or storage tanks located in the vicinity of producing pads. 

In order to model methane emissions linked to these upstream operations, two types of data will be used: methane emissions and oil \& gas production. 

\subsection{Data Sets}

\subsubsection{Methane Emissions Data}

Kayrros methane emissions data are focused on upstream oil \& gas operations. These emissions are visible from space using hyperspectral satellite imagery (from e.g. the Sentinel-5P satellite, part of the European Space Agency’s (``ESA") Copernicus constellation), which identifies high concentrations of methane above major production areas like the Permian basin in the US or Iraq. To model total anthropogenic emissions in a production area from these methane concentration grids, we use a “full-inversion model” as in \cite{fullInversion1} \cite{fullInversion2} \cite{fullInversion3}. Our full-inversion model uses satellite images (filtered to distinguish anthropogenic emissions from natural methane concentrations in the atmosphere), atmospheric simulations and on-ground operational data (well completions, flaring intensity, etc.) to obtain an accurate model of total emissions in a production area. 

The second data source for methane emissions is a public database: the IEA’s Global Methane Tracker 2023 \cite{ieaGMT23}. In its documentation, the IEA details that its estimated emissions are obtained by first computing US emission intensities based on measurement studies and satellite readings. For other countries, the emissions are determined by scaling US emissions intensities. Various country-specific data sets are used in the scaling process, including infrastructure age, types of operators in the country, average intensity of flaring as well as governance indicators such as the strength of regulation and oversight, government effectiveness and regulatory quality in the country.
Country-level methane emissions estimates are then obtained by applying the country emissions intensity to the activity data. When some other robust data source is available, some adjustments can be made to the scaling factors. Overall, while IEA methane emissions take into account a lot of parameters, they remain estimates rather than measurements.

Kayrros measurements and IEA estimates are combined based on a ``best available data" approach, giving priority to measurements rather than estimates. This enables an allocation of methane emissions to basins, countries or regions and Kayrros measurements bring an additional granularity to the allocation of emissions with measurements on basins when only country-level allocation is possible with the IEA estimates.

\subsubsection{Oil \& Gas production data}

In order to estimate the methane intensity of a country, the oil and gas production of this country is required. For crude oil (and other liquids such as lease condensate and NGLs) production in 2022, our source is the US Energy Information Administration (``EIA") international database for petroleum and other liquids \cite{eiaOil}. This provides detailed information for each type of liquid for a large number of countries.

Regarding natural gas production, an extrapolation on 2021 data is used as there is no reliable public database that provides data for all countries in 2022. The EIA international database for natural gas only includes information for 2021 \cite{eiaGas} and the JODI Gas World Database provides information for a few countries in 2021 and 2022 \cite{jodiGas}. The coverage being more extensive with the EIA database, we use it as a reference for gas production. To get an estimate of the 2022 natural gas production by country, the following extrapolation approach is used: 
\begin{itemize}
    \item when JODI data is available for both 2021 and 2022 for a specific country, the observed variation between these years is applied to the 2021 production number of the EIA and
    \item if JODI data is unavailable for these years, the 2022 production is estimated based on the trends in the historical data of the EIA database for the country observed.
\end{itemize}

\subsection{Methodology for modeling methane emissions}

Given the data detailed above, the following methodology will be used to model emissions linked to upstream activities with different regional footprints. For the sake of relevance, these production profiles will be modelled after existing companies.

The first step of the methodology is to build the model production profile, hereafter called {\em Model} company. A production profile is a list of regions in which the modelled company is known to have upstream operations associated with a production quantity. Company reports and publicly available data are used to build these model production profiles.

For each of the regions included in the Model, a methane intensity is computed. It is considered here as the methane emissions of the region, as obtained with the ``best available data" approach previously described, divided by the global oil and gas production of the region. 
With this method, the methane intensity of the production in a specific region is considered uniform over the whole regional production and does not take into account any actions of individual operators that would justify considering different methane intensities within a single region.

With the Model built and the methane intensities in each region included, regional emissions of the Model company can be computed by multiplying the methane intensity of the region by the production considered in the region for the Model. The total emissions of the Model company are the result of the aggregation of all regional emissions. This metric then gives a model of the methane intensity of the profile when divided by the relevant production.

The methane emissions and intensity obtained for the Model company can be compared to the reported metrics of the company after which the production profile was modelled. This enables the creation of a benchmark for methane emissions and methane intensity for different upstream production profiles and companies.

\subsection{Assumptions}

This methodology makes a number of assumptions to get a reliable idea of the methane emissions of a Model company.

\begin{itemize}
    \item As mentioned before, the methane intensity of the production in a specific region is considered uniform over the whole regional production and does not take into account any actions by individual operators that would allow them to reduce their emissions below the country average. Here, a country/basin is the finest granularity level available in terms of oil and gas production as well as methane emissions.
    \item Kayrros data covers basins in specific countries. These basins do not always cover the whole country, however, the assumption here is that the methane intensity in the basins is representative of the situation in the whole country. For example, in the US, three basins are covered by Kayrros data: the Permian, Anadarko and Appalachian basins. Each of these basins has a specific methane intensity. The methane intensity considered for the rest of the US is the average of these methane intensities, weighted by the oil and gas production in each basin.
    \item For some companies, the country-level production data is not always available for every country the company operates in. The aggregation of the countries for which the split is not available is then considered as the highest level of granularity available and a new methane intensity is computed for this aggregation based on the total production and emissions of the aggregation, considered uniform over all the countries within.
    \item Some countries are not included in the IEA’s Global Methane Tracker 2023. However, over the scope of the production profiles studied here, the production in these countries is negligible. The methane intensity of the Model is computed with the production from covered countries. Methane emissions can then be modelled either for the full production, considering that the missing production would not have an impact on the methane intensity of the Model, or only for the production in covered countries.
    \item Model methane intensities are obtained from different sources: Kayrros data and IEA estimates. Each source does not cover the exact same scope: Kayrros data covers all upstream emissions (including emissions linked to super-emitters \cite{doi:10.1126/science.abj4351}) and IEA estimates also cover upstream but do not include super-emitters.
\end{itemize}

\section{Results}

In this section we present the anonymised results for 42 oil and gas companies, representing around 34\% of the global oil and gas production, using the methodology presented above. These companies are categorized based on the following definitions in order to allow relevant comparisons:
\begin{itemize}
    \item NOC (National oil company): National operator, controlled by the state (can have investments outside of its national frontiers)
    \item Integrated oil and gas company: Companies that engage in the exploration, production, refinement and distribution of oil and gas
    \item Independent oil company: other company, operating abroad or in several countries on the upstream segment.
\end{itemize}


The list of 42 companies split by category is presented in Table \ref{tab:companies}.


\begin{table}
\centering
\caption{\label{tab:companies}Companies by category}
\begin{tabular}{c|c|c}
\textbf{National Oil Companies} & \textbf{Integrated Oil Companies} & \textbf{Independent Oil Companies}\\ \hline
Aramco & BP & APA Corporation\\
Ecopetrol & CEPSA & ConocoPhillips\\
Gazprom & Chevron & Devon Energy\\
Pemex & ENI & DNO ASA\\
Petrobras & Equinor & EOG Resources\\
Petronas & ExxonMobil & Genel Energy\\
QatarEnergy & Inpex & Gulf Keystone Petroleum\\
Qatar Gas & Lukoil & Hess\\
Sonatrach & Oxy & Marathon Oil\\
TAQA & OMV & Murphy Oil Corporation\\
& Repsol & Novatek\\
& Shaanxi Yanchang Petroleum & Perenco\\
& Shell & Pioneer Natural Resources\\
& Suncor & ShaMaran Petroleum\\
& Surgutneftegas & Wintershall Dea\\
& Tatneft &\\
& TotalEnergies &\\
\end{tabular}
\end{table}

\subsection{Model methane intensity}

\begin{figure}
    \caption{\label{fig:distrib_mi}Distribution of model methane intensity}
    \centering
    \begin{minipage}{0.48\textwidth}
        \centering
        \includegraphics[width=1\textwidth]{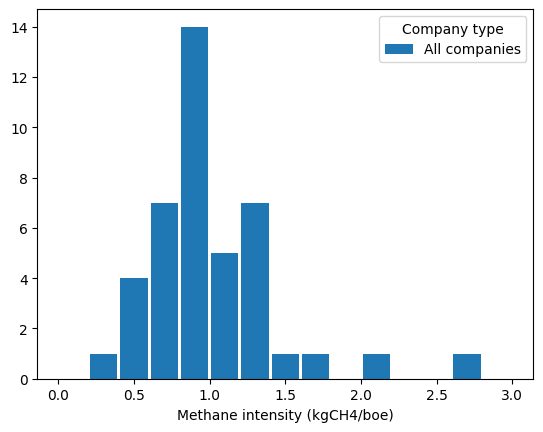} 
    \end{minipage}\hfill
    \begin{minipage}{0.48\textwidth}
        \centering
        \includegraphics[width=1\textwidth]{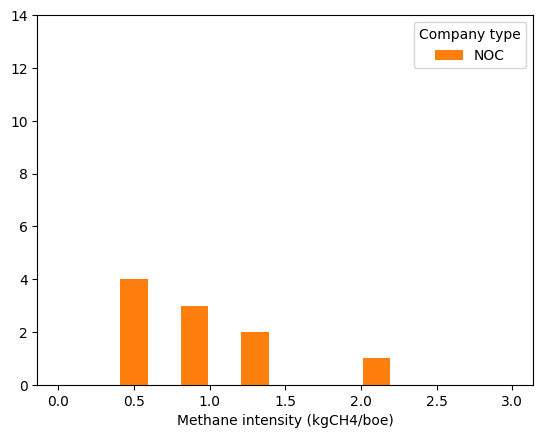} 
    \end{minipage}\hfill
    \begin{minipage}{0.48\textwidth}
        \centering
        \includegraphics[width=1\textwidth]{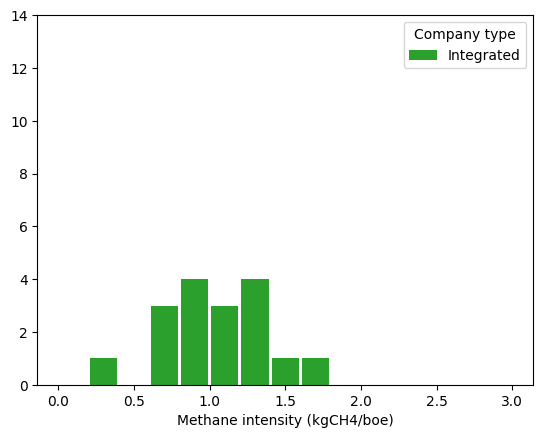} 
    \end{minipage}\hfill
    \begin{minipage}{0.48\textwidth}
        \centering
        \includegraphics[width=1\textwidth]{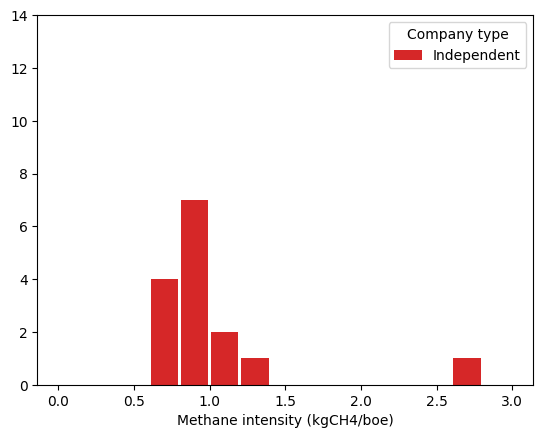}
    \end{minipage}\hfill
\end{figure}

Over these companies, the methane intensities modelled ranged from 0.32 to 2.75 kgCH4/boe, with an average of 1.0 kgCH4/boe and a median of 0.88 kgCH4/boe. Figure \ref{fig:distrib_mi} shows a distribution of the methane intensity close to a Gaussian distribution around 0.8 - 1.0 kgCH4/boe, with two companies modelled at more than 2.0 kgCH4/boe. These two models are based on a NOC and an Independent company respectively.

When breaking down companies by type, we can see that all company model groups have their methane intensities centred around 0.8 - 1.0 kgCH4/boe.



\begin{figure}
\centering
\caption{\label{fig:mi_boxplot}Model methane intensity by type of company}
\includegraphics[width=0.75\linewidth]{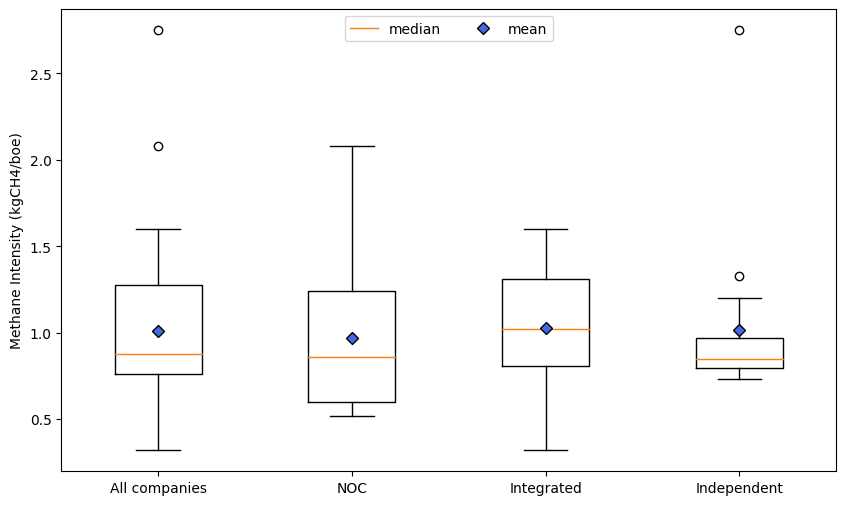}
\end{figure}

Figure \ref{fig:mi_boxplot} shows the distributional characteristics of the different company types regarding methane intensity models. The model set presented here has a global average of 1.0 kgCH4/boe. The average for Integrated and Independent is slightly higher than this number, with 1.03 and 1.02 kgCH4/boe respectively. This is offset by the average methane intensity of the NOC group that is at 0.97 kgCH4/boe.

The relative standard deviation of the values for NOCs and Independent companies shows that the region of operation has a high impact on the methane intensity of the production profile considered, with values of 50.8\% and 50.0\% respectively, whereas Integrated companies have relative standard deviations of 33.9\%.

\subsection{Reported methane intensity}

\begin{figure}
    \centering
    \caption{\label{fig:reported_distrib_mi}Distribution of company reported methane intensity}
    \begin{minipage}{0.48\textwidth}
        \centering
        \includegraphics[width=1\textwidth]{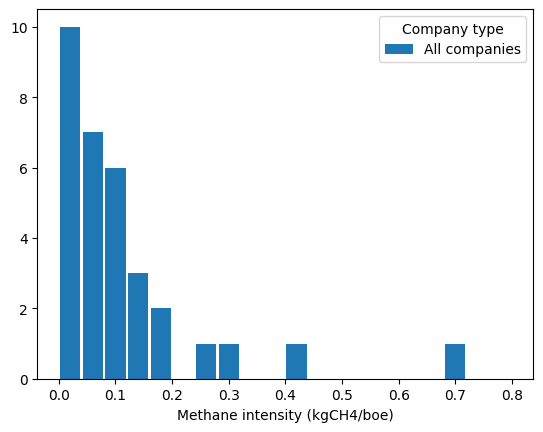} 
    \end{minipage}\hfill
    \begin{minipage}{0.48\textwidth}
        \centering
        \includegraphics[width=1\textwidth]{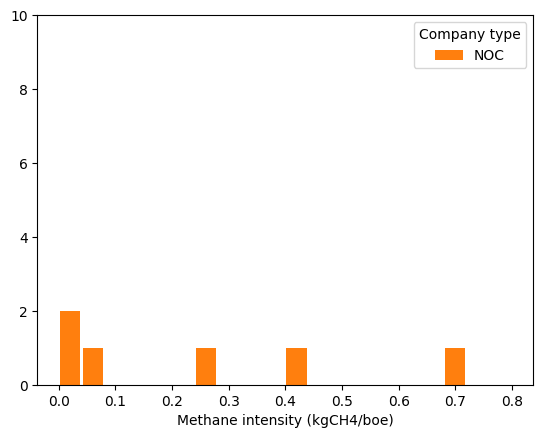} 
    \end{minipage}\hfill
    \begin{minipage}{0.48\textwidth}
        \centering
        \includegraphics[width=1\textwidth]{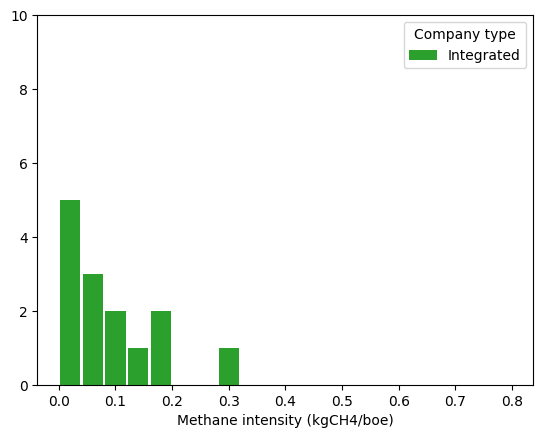} 
    \end{minipage}\hfill
    \begin{minipage}{0.48\textwidth}
        \centering
        \includegraphics[width=1\textwidth]{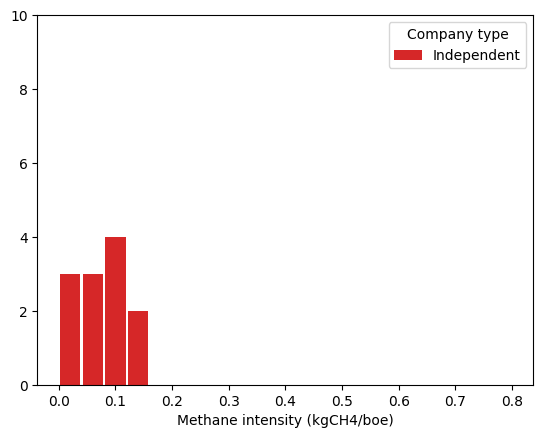}
    \end{minipage}\hfill
\end{figure}

\begin{figure}
\centering
\caption{\label{fig:reported_boxplot}Reported methane intensity by type of company}
\includegraphics[width=0.75\linewidth]{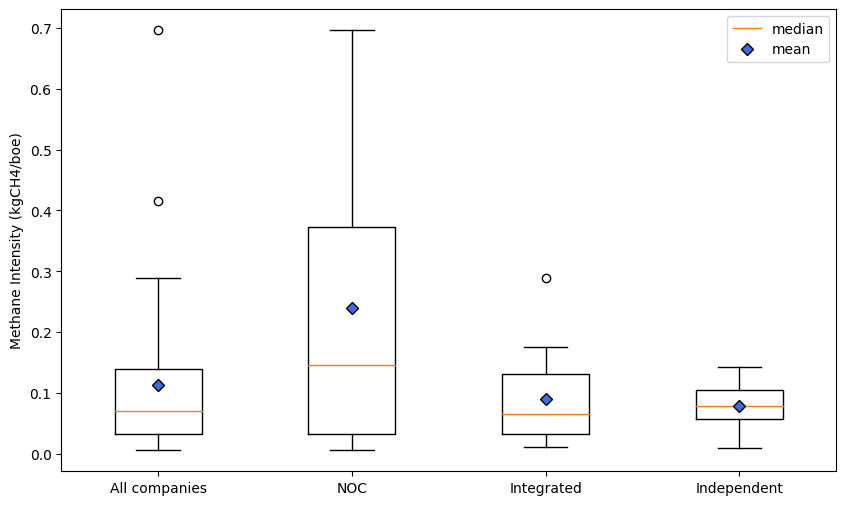}
\end{figure}

Out of the 42 companies considered in this study, 32 published a report on their methane emissions: 6 NOCs, 14 Integrated and 12 Independent. The distribution of the corresponding methane intensities is shown in Figure \ref{fig:reported_distrib_mi} and the distributional characteristics are in Figure \ref{fig:reported_boxplot}.

The relative dispersion of the reported methane intensities is much higher than for the estimated results, especially for NOCs and Integrated companies that show a relative standard deviation of 115\% and 86.6\% respectively. Independent companies are still relatively dispersed with 53.4\% relative standard deviation. 

\subsection{Difference between modelled and reported methane intensities}
The emissions of the models presented above can be compared to the methane emissions reported by the company on which the model is based, in its sustainability reporting or its public communication. Here, we will use the ratio between model methane intensity and reported methane intensity, in order to present the results for our set of production profiles for reporting companies.

\begin{figure}
\centering
\caption{\label{fig:factor_boxplot}Ratio between modeled and reported intensities by type of company}
\includegraphics[width=0.75\linewidth]{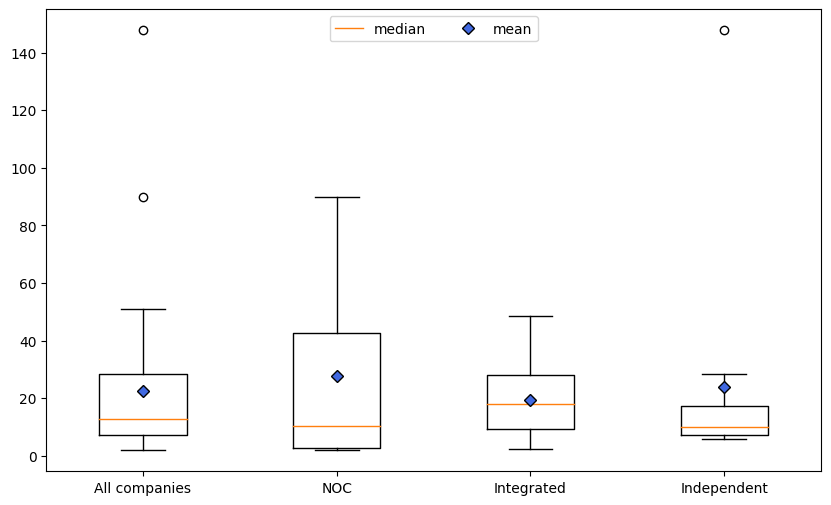}
\end{figure}

\begin{figure}
\centering
\caption{\label{fig:factor_boxplot_wo_outliers}Factor between modelled and reported metrics by type of company (without outliers)}
\includegraphics[width=0.75\linewidth]{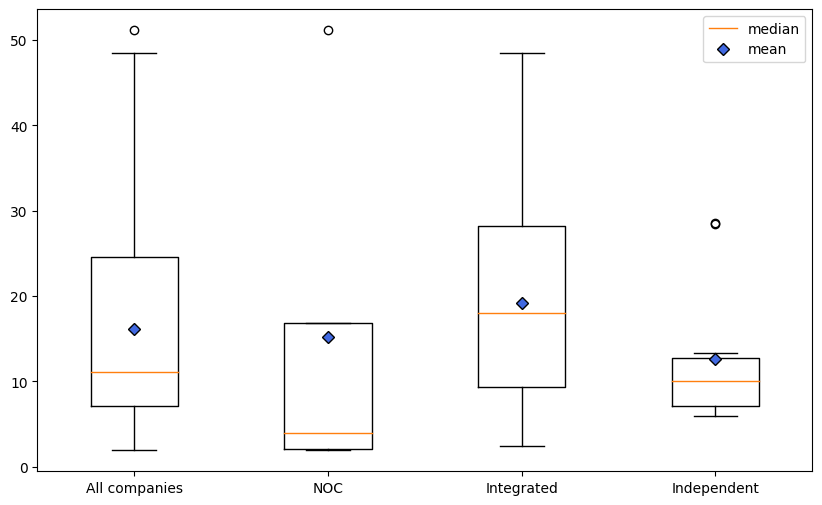}
\end{figure}

The ratio distribution of the different company types is shown in Figure \ref{fig:factor_boxplot}. An outlier in the NOCs and one in the Independent oil companies are strongly affecting the results presented and will not be included in the metrics presented hereafter and in Figure \ref{fig:factor_boxplot_wo_outliers}.



The global average ratio on our set of companies in Figure \ref{fig:factor_boxplot_wo_outliers} is 16.1, with a relative standard deviation of 79.3\%. Each type of company within this set does not contribute similarly to these global metrics. Integrated companies have the highest average ratio between model and reporting out of all the types of companies at 19.2 and a relative standard deviation of 64.7\%, which means that some Integrated companies reporting are much closer to the values for their corresponding model than others.

NOCs have the lowest median ratio with 4.0 but their average is the second highest at 15.2, strongly influenced by one of the companies in the category that has a factor of 51.2 while the remaining 4 have an average of 6.2. Independent companies have the lowest average with 12.6 and a relative standard deviation of 65.2\%, which once again shows the importance of the regions of operation in the results of different models.

\section{Discussion}

The results presented in this paper highlight the impact of the region of production on modelled methane intensities. Factors such as extraction methods, infrastructure quality, and regulatory frameworks can vary significantly from one region to another, leading to differing methane emission patterns and the different combinations of production footprints can vary significantly from one profile to another.

The issue of potential under-reporting by companies has emerged as a concern in our investigation of methane emissions. As highlighted in an article from The Washington Post \cite{theWPost}, there are indications that some companies may not be fully transparent or accurate in reporting their methane emissions. This lack of comprehensive reporting could be attributed to various factors, such as limited monitoring capabilities, inadequate regulatory oversight, or a lack of incentives for companies to disclose accurate data.

The measurements conducted by Kayrros exhibit an average relative uncertainty of 20\%, which can be attributed to various contributing factors. These factors notably encompass sensor precision, the inherent uncertainty associated with the methodology employed for background computation, uncertainties arising from meteorological data, and considerations related to the transport model. The combined influence of these factors may have a substantial impact on the resulting outcomes. Acknowledging and quantifying uncertainties is essential in interpreting and utilizing methane emission data responsibly. As we continue to refine measurement techniques and expand data collection efforts, we can aim to reduce the level of uncertainty associated with methane emission measurements.

Assuming the reliability of global data,  the primary challenge lies in dealing with companies that do not report their emissions. It's crucial to recognize that this subset of companies could potentially introduce a significant bias into our statistical analyses. On the other hand, it's important to note that we have focused our analysis on approximately 16\% of the global oil and gas production which includes the largest producers of the industry. However, this limited coverage may introduce additional nuances to the interpretation of our analysis results. Therefore, while our findings provide valuable insights into the subset of companies analyzed, they may not be fully representative of the entire industry. Promoting environmental transparency and the standardization of reporting norms remains crucial for enhancing data quality and mitigating potential biases in future analyses.

Finally, the use of the Sentinel-5 Precursor satellite from the Copernicus program for studying methane emissions has both advantages and disadvantages. First, the availability of open data enhances scientific collaboration and large-scale analysis. Moreover, its daily coverage of the entire planet provides exceptional global monitoring of methane emissions. The required computational time is limited, making data processing more manageable. However, it's important to note that other satellites offer higher spatial resolution, which can be crucial for studies requiring fine local-scale precision. Furthermore, some satellite imaging technologies allow for the correction of water-related disturbances, which can be a significant challenge in methane emissions data acquisition in offshore areas. These considerations pave the way for new studies that could combine the advantages of Sentinel-5 Precursor with other data sources for a more comprehensive understanding of methane emissions using an inversion framework.


\section{Conclusion}

The oil and gas industry is the largest energy sector source of methane emissions, which can be reduced cost-effectively. One of the major obstacles to this reduction is correctly quantifying methane emissions and regularly following the impact of the efforts made. Satellite-based measurements and the potential of future measurements with remote sensing technology could help tackle this issue.

In this paper, we utilize satellite measurements and methane emission modelling to assess the influence of various model companies' operational regions on their emissions. By modelling the production profiles of existing companies, we can compare our metrics with actual reporting by these companies.

This scientific paper makes a significant contribution compared to existing inventories by adopting a systematic and comprehensive top-down approach, utilizing satellite imagery whenever possible. This innovative method allows for a more comprehensive modelling of company-level methane emissions, providing a holistic perspective that can be put in front of  the reported environmental activities of the studied companies. The incorporation of satellite imagery into the analysis process adds an independent measurement and valuable objectivity, resulting in more reliable and detailed data on upstream oil and gas emissions. Thus, this paper expands our methodological toolkit for modelling the environmental impact of companies, paving the way for further research and more informed decision-making in sustainability and environmental policy.

\section{Going further}

As research on methane emissions continues to evolve, there are several axes that could improve our methane emissions model. The pursuit of these improvements is crucial as it will lead to more effective strategies for mitigating methane's impact on climate change. Here, we outline three key areas that offer promising prospects for refining our methane emission model.

One potential area of improvement is increasing the number of measurements by integrating future high-resolution satellite data with advanced remote sensing techniques to identify and quantify more precisely methane emissions. These new technologies could also enable the expansion of the spatial coverage of full-inversion models and give a more comprehensive global perspective on measured methane emissions.

Another improvement could come from the consideration of super-emitters, which are critical targets for emission reduction efforts. Integrating methane emissions from these super-emitters into estimates from the IEA can provide a more complete picture of the sources driving methane emissions.

A third axis for improving our methane emission model is to achieve better regional granularity by using other sources of information with better spatial resolution on both emissions and production estimates.

\section*{Acknowledgement}
I would like to thank Amine Baccar for his contributions and support throughout the research process, which enriched the quality of this paper.

\clearpage

\bibliographystyle{ieeetr}
\bibliography{refs}

\end{document}